\documentclass[prb,twocolumn,floatfix,showpacs]{revtex4-1}
\usepackage{epsfig,subfigure,amssymb,amscd,amsmath,graphicx,amsfonts,mathtools}
\usepackage{times,longtable,setspace,fancyhdr,flushend}

\newcommand{\non}{\nonumber}

\newcommand{\bea}{\begin{eqnarray}}
\newcommand{\eea}{\end{eqnarray}}
\newcommand{\be}{\begin{equation}}
\newcommand{\ee}{\end{equation}}
\newcommand{\ba}{\begin{align}}
\newcommand{\ea}{\end{align}}
\newcommand{\braket}[2]{\langle #1|#2\rangle}
\newcommand{\ket}[1]{     |    \,    #1    \rangle}
\newcommand{\bra}[1]{  \langle #1  \,  |}

\newcommand{\phd}{ {\phantom\dagger} }

\newcommand{\lr}{\leftrightarrow}
\newcommand{\ud}{\updownarrow}

\usepackage[usenames]{color}

\begin{document}

\title{Many-body Majorana operators and the equivalence of parity sectors}

\author{G. Kells}
\affiliation{ Dublin Institute for Advanced  Studies, School of Theoretical Physics, 10 Burlington Rd, Dublin, Ireland. \\ Department of Mathematical Physics, Maynooth University, Maynooth, Co. Kildare, Ireland.}

\begin{abstract}
The one-dimensional p-wave topological superconductor model with open-boundary conditions is examined in its topological phase.  Using the eigenbasis of the non-interacting system I show that, provided the interactions are local and do not result in a closing of the gap, then even and odd parity sectors are unitarily equivalent. Following on from this, it is possible to define two many-body operators that connect each state in one sector with a degenerate counterpart in the sector with opposite parity. This result applies to all states in the system and therefore establishes, for a long enough wire, that all even-odd eigenpairs remain essentially degenerate in the presence of local interactions. Building on this observation I then set out a full definition of the related many-body Majorana operators and point out that their structure cannot be fully revealed using cross-correlation data obtained from the ground state manifold alone. Although all results are formulated in the context of the 1-dimensional p-wave model, I argue why they should also apply to more realistic realisations (e.g. the multi-channel p-wave wire and proximity coupled models) of topological superconductivity.
\end{abstract}


\date{\today} \maketitle
\thispagestyle{fancy}

Since the realisation that topological superconductors can support Majorana bound states, \cite{Read2000,Kitaev2001} significant advances have been made towards generating the necessary effective  p-wave symmetry.  There are now a large number of candidate systems in which these Majorana states could  potentially be observed, \cite{Alicea2012,Beenakker2013,Stanescu2013} the most well-known being those based on proximity-coupled semiconductor nano-wires. \cite{Oreg2010,Lutchyn2010} In these nano-wire systems, observations of  anomalous zero-bias conductances are a strong experimental indication of the Majorana modes,\cite{Mourik2012,Das2012,Churchill2013} although they are not yet fully conclusive.\cite{Liu2012, Pikulin2012, Bagrets2012, Kells2012, Lee2012, Lee2013}   Recently, approaches using magnetic molecules, whose bound states can be resolved both energetically and spatially, have also attracted  considerable interest, see e.g. Ref. \onlinecite{Nadj-Perge2014} and references within.  Much of the excitement surrounding topological superconductors comes from the knowledge that with each each pair of Majorana zero-modes one can associate an effective ground-state degeneracy, within which it should be possible to manipulate quantum information robustly using non-local braiding operations. \cite{Ivanov2001,Nayak2008, Alicea2011}

A topological ground-state degeneracy is a key signature of what is known as strong topological-order. \cite{Wen2004} Although now it forms one element of a growing literature on interacting topological signatures, see for example Refs. \onlinecite{Fidkowski2010,Turner2011,Fidkowski2011b,Gurarie2011,Manmana2012,Meidan2014}, its enduring usefulness stems from its direct applicability to both free-fermion and interacting many-body systems.   An interesting feature of the degeneracy associated with localised zero-energy Majorana excitations is that they are formulated using solvable quadratic Hamiltonians  and therefore their existence implies that every eigenstate of the system, and not just the groundstate, has an eigen-partner of opposite parity at the approximately the same energy.  Thus the low energy excitation spectrum is governed in these case by \cite{Kitaev2001}
\be
\label{eq:H1}
H_{\text{eff}}=i t \gamma_L \gamma_R
\ee
where $t=a \exp(-L_x/\xi)$ where $\gamma_L$ and $\gamma_R$ are the Majorana quasi-particles localised at both ends of the wire. This  effective picture is valid when applied to the ground state manifold or in a weakly interacting regime such that a treatment on the level of mean-field theory is accurate.  Beyond the mean-field description however one expects that this quasi-particle description breaks down and indeed the low-energy single particle excitation spectrum \label{eq:Heff1} is actually a special case of the more complicated sum
\bea
\label{eq:Heff2}
H_{\text{eff}}=   \sum_{n } t_n (\ket{n_e} \bra{n_e}-\ket{n_o} \bra{n_o})
\eea
where the $t_n$ do not necessarily have to be the same.  

In this paper however I show that, in the topological phase of an infinite open wire, provided the interactions are local and do not result in a closing of the gap, even and odd parity sectors of the p-wave Hamiltonian are spectrally equivalent. Remarkably this argument holds also in the limit of strong interactions and thus implies that in the topological phase all $t_n \rightarrow 0$ as the wire length is increased and thus the effective description (\ref{eq:H1})  again becomes valid  simply because all coupling constants have become exponentially small. As a direct consequence of this, it then becomes possible to define infinite lifetime zero-energy excitations that persist even in the presence of strong interactions.

The nature of the Majorana states has been addressed previously in the context of single interacting wires using bosonization\cite{Gangadharaiah2011,Stoudenmire2011,Fidkowski2011, Lutchyn2011,Sela2011,Lobos2012} and additional numerical approaches related to the Density Matrix Renormalization Group. \cite{Stoudenmire2011,Thomale2013}  Identical microscopic models also arise in the context of  nano-wire bridges on superconducting islands. \cite{Hassler2012}  Recently the robust nature of the Majorana degneracy  to hopping disorder was contrasted to the general instability of zero-modes in  wider para-fermionic family of 1d models. \cite{Jermyn2014}   Additional contributions to this general area also include examining the occurrence of zero-modes in interacting Hamiltonians with odd numbers of Majorana particles. \cite{Akhmerov2010,Goldstein2012,Yang2014}   This work presented here compliments the aforementioned approaches by establishing that the Majorana degeneracy  applies  to all eigenstates of the interacting model and thus allows the straightforward definition of many-body Majorana operators that are well defined quasi-particle excitation of the interacting system.    Moreover, the general arguments presented here do not rely on the assumption of an odd number of participating Majorana modes, a condition which necessarily implies the existence elsewhere of a non-participating/unpaired Majorana zero-mode. 

\vspace{2mm}
This paper examines the combined Hamiltonian
\be
H=H_0 + H_{I}.
\ee
where $H_0$ is the 1D $p$-wave superconducting model \cite{Kitaev2001}  and $H_{I}$ is an electron-electron interacting term.  The bare tight-binding Hamiltonian for a single wire is given by
\begin{eqnarray} \label{Hmicro}
  H_0 & = & - \mu \sum_{x=1}^{N_x} c^{\dagger}_{x} c^{\phantom \dagger}_{x} \\
 \ & \ & - \sum_{x=1}^{N_x-1} \left( t c^{\dagger}_{x} c^{\phantom \dagger}_{x+1} +|\Delta| e^{i\phi} c^{\dagger}_{x} c^{\dagger}_{x+1} + \mbox{h.c.} \right), \nonumber
 \label{eq:H0} 
\end{eqnarray}
 where $\mu$ is a chemical potential, $t$ the hopping energy, $|\Delta|$ the magnitude of the pairing potential and $\phi$ the superconducting phase. \cite{Continuum}  The general form of the interaction term can be written as
\be
\label{eq:Hint}
H_{I} = \frac{1}{4}  \sum_{x_1 x_2 x_3 x_4}  \bar{\nu}_{x_1 x_2 x_3 x_4}  c^\dagger_{x_1}  c^\phd_{x_3}   c^\dagger_{x_2}  c^\phd_{x_4}  
\ee
where $\bar{\nu}_{x_1 x_2 x_3 x_4} = \nu_{x_1 x_2 x_3 x_4}-\nu_{x_1 x_2 x_4 x_3}$.  In the topological p-wave wire literature one often finds the specific form  $\nu_{x_1 x_2 x_3 x_4} = I (x_1,x_2) \delta_{x_1,x_3} \delta_{x_2,x_4}$ with $x_2 = x_1+1$.The subsequent analysis however can be applied to the more general form above, provided we insist that the term is local.

The Hamiltonian $H_0$  may be written in terms of free fermions $H_0=\sum_n \epsilon_n (\beta_n^\dagger \beta_n^{\phantom \dagger} - 1/2)$ by a Bogoliubov transformation  
\bea
\label{eq:beta}
c_x^\dagger &=& \sum U_{xn}^* \beta_n^\dagger + V_{xn} \beta_n \\
c_x^{\phantom \dagger} &=& \sum U_{xn} \beta_n^{\phantom \dagger} + V_{xn}^* \beta^\dagger_n \non
\eea
where, without loss of generality, we can choose the phase $\phi=0$ such that $U$ and $V$ are real \cite{Kitaev2001}.

When $|\Delta|>0$ and $|\mu| < 2t$ the $H_0$ system is known to be in a topological phase with a Majorana zero modes exponentially localized at each end of the wire\cite{Kitaev2001}. In the limit $N_x\rightarrow \infty$ the (L)eft and (R)ight Majorana modes have precisely the energy $E=0$ and the corresponding operators have the form
\bea \label{eq:Majdef}
\gamma_L  &=&  i \sum_x  (c^\dagger_x - c_x) u_L(x) = i (\beta_1^\dagger - \beta_1^{\phd}) \\
\gamma_R  &=&  \phantom{i} \sum_x  (c^\dagger_x + c_x) u_R(x) = \phantom{i} (\beta_1^\dagger + \beta_1^{\phd}) \non 
\eea
Inverting \eqref{eq:Majdef} we can write the complex fermion zero-mode responsible for the degeneracy as
\be
\beta_{1}^\dagger = \frac{1}{2} (\gamma_R -i  \gamma_L) \quad \beta_{1}^\phd= \frac{1}{2} (\gamma_R + i \gamma_L)
\ee 
For hard-wall boundary conditions one finds that
 \bea
u_L(x) &=& C A^x \sin( \theta x) \\ 
u_R(x)  &=& C  A^{\bar{x}} \sin (\theta \bar{x})
\eea
where $C$ is a normalisation factor, $\bar{x}=N_x-x$, and 
\be
A = \sqrt{\frac{t-|\Delta|}{t+|\Delta|}}, \quad \theta =\cos^{-1} (\frac{-\mu+2t }{2 \sqrt{t^2-|\Delta|^2}}). \nonumber
\ee
The Majorana wave-functions in this case are therefore oscillating functions inside a exponentially decaying envelope. The correlation length is given by $\xi=t/\Delta$. 

Of course these exact expressions for Majorana wavefunctions are only strictly true in the infinite smooth wire. Although the precise local character of the wave functions may change if we introduce for example disorder,  we will still have well defined zero-modes provided that the functions $u_L$ and $u_R$ decay exponentially.  In what follows we will find it useful to distinguish between coordinates at the left of the system $x_L$ and coordinates on the right of the system $x_R$. What actually constitutes the left and right (or middle) is determined by the coherence length $\xi$ and the length $N_x$ but by allowing ourselves the freedom to increase the wire length we can always assume that  $u_R(x_L) \rightarrow 0 $ and $u_L(x_R) \rightarrow 0$.  In terms of the (now real) matrices $U$ and $V$, because
\bea
\label{eq:UVdecomp}
U_{x,1}&=& u_R(x) + u_L(x) \\  \non V_{x,1}&=& u_R(x) -u_L(x),
\eea
we have \bea
\label{eq:UVdecomp2}
U_{x_L,1} &=& \phantom- u_L(x_L), \quad U_{x_R,1}= u_R(x_R), \\ \non
V_{x_L,1} &=& - u_L(x_L), \quad V_{x_R,1}= u_R(x_R). 
\eea

Before addressing the interactions directly we need to address briefly our book-keeping of bulk excitations of the system. Although expressions for these excitations are also possible to write down, they can be complicated and knowing their precise form is not necessary for what we want to show.  What is important is that that all eigenstates of the $H_0$ system can be written in terms of these $\beta^\dagger_n$ operators acting on the ground state.  For an 8 site system in Fock space we would have for example $\ket{10000000} = \beta_1^\dagger \ket{00000000}$
where $\ket{00000000}$ is the ground state defined as
\be
\ket{00000000} = \mathcal{N} \prod \beta_{n} \ket{\text{ref}}
\ee
such that $\beta_n \ket{00000000} =0$ for all $n$.  The factor  $\mathcal{N}$ is a normalisation constant and $\ket{\text{ref}}$ is a reference state, often chosen to be the vacuum of the $c$-fermions defined above i.e. $ c_x \ket{\text{ref} } =0$  for all sites $x$.

In the case that $\beta_{1}^\dagger$ is  a zero mode ($\epsilon_1 \rightarrow 0$), any two states 
\be
 \ket{0abcd.....} \quad \text{and} \quad  \ket{1abcd ....} 
\ee
will have approximately the same energy.  In this case one should note of course that within a large system the bulk state energies become arbitrarily close to one another and thus when speaking of the so called Majorana degeneracy for higher energy states of the system we simply mean that eigenstates  with a different occupancy of the zero mode. For the two such lowest energy states in the system the first index also indicates the parity, although this is not always the case. For example the state $\ket{11000000}$ has even fermion parity but has the zero energy mode occupied.  It will therefore be important to distinguish between eigenstates in two different ways.  In the first we simply  denote  $\ket{n}_e$ for total even occupied states and  $\ket{n}_o$ for its counterpart in the odd sector with the opposite occupancy on the zero energy mode.  In what follows however it will also be helpful to indicate the occupation of the zero mode explicitly and define $\ket{n}_0 $ as the states with the mode empty and $\ket{n}_1$ as states with the mode occupied.  A useful sub-division of this latter labelling is one which defines the total number of fermionic occupations $N$, not including the zero mode $\beta_1^\dagger \beta_1$ and a sub-label $j$ denoting the ${N_x-1 \choose N}$ different possibilities with this set. In this case if we set $n = j +\sum_{i=0}^{N-1} {N_x-1 \choose i}$ we relate these two labelling schemes:
\bea
\ket{n}_e &=& \ket{j,N}_0 \;\; \text{and}\;\; \ket{n}_o = \ket{j,N}_1  \quad \text{when N is even} \non, \\
\ket{n}_e &=& \ket{j,N}_1 \;\; \text{and} \;\;\ket{n}_o = \ket{j,N}_0  \quad \text{when N is odd} \non.
\eea

Now we are in a position to show that the weak interaction term (i.e. as long as it does not close the gap and trigger a quantum phase transition to a non-topological phase) does not destroy the equivalence between even and odd sectors.  Let us first expand $H_{\text{I}}$ in the eigenbasis of $H_0$.  Substituting (\ref{eq:beta})  into \eqref{eq:Hint} we get
\bea
\label{eq:H_expand}
H_{\text{I}} &=&   \frac{1}{4} \sum_{x_1 x_2 x_3 x_4} \nu_{x_1 x_2 x_3 x_4} \times  \\ && \sum_i (U_{x_{1} i}^* \beta_i^\dagger + V_{x_{1} i}^{\phantom*}  \beta_i^{\phd})   \sum_j (U_{x_{3} j}^{\phantom*} \beta_j^{\phd }+ V_{x_{3} j}^* \beta_j^\dagger ) \non \\
&& \sum_k (U_{x_{2} k}^* \beta_k^\dagger + V_{x_{2} k}^{\phantom*}  \beta_k^{\phd} ) \sum_l(U_{x_{4} l}^{\phantom*} \beta_l^{\phd} + V_{x_{4} l}^* \beta_l^\dagger ) \quad  \non
\eea
Although it is a technical exercise in indexing and sign counting, it is a computationally simple task to calculate any matrix element  $\bra{n} H_{I} \ket{m}$ of this interacting term.  An important observation regarding terms of $H_I$, and indeed any parity preserving operator in the $H_0$ eigenbasis, is that for each term that applies to one parity sector there is counterpart in the other sector which we can obtain by switching  occurrences of  $\beta_1$ with coefficient $U_{x,1}$ or $V_{x,1}$ with the $\beta_1^\dagger$  occurring in the same product in the expansion but with swapped and negated coefficients $- V_{x,1}$ or $-U_{x,1}$. 

The equivalence between even and odd sectors is proven by showing that the effective Hamiltonians describing the energies of bands  $\ket{j,N}_0$ and  $\ket{j,N}_1$ are the same to an order of perturbation theory that scales with the length of the system.  To make this argument we first show that, in the topological phase, matrix elements between states within the same band , of any parity preserving operator that is local in position space  ( $O_{local}$) ,   are the same in both even and odd sectors, i.e.
\be
\phantom|_0 \bra{j,N} O_{local} \ket{k,N} _0= \phantom|_1 \bra{j,N} O_{local} \ket{k,N} _1
\label{eq:diag_equiv}
\ee
From here we can then argue that a degenerate perturbative expansion of energies about the special point with flat dispersion is convergent for a finite range of perturbing parameters and therefore that, for a long enough wire, both even and odd sector Hamiltonians are  equivalent.  

The reason (\ref{eq:diag_equiv}) is true is that while the even and odd sub-Hamiltonians ($H_{(e)}$ and $H_{(o)}$) are not generally equal, they only differ in those matrix elements which are connected by terms containing a single unpaired $\beta^\dagger_1$ or $\beta_1$.  These terms (denoted D) carry coefficients ( either $U_{x,1}$ or $V_{x,1}$ depending on where in the expansion they occured) which factor into left $\pm$ right superpositions 
\bea
\label{eq:LRdecomp2}
\phantom|_p \bra{j,N} D  \ket{k,M}_p  = L_{nm} \pm (-1)^p R_{nm} 
\eea
and are therefore generally not equivalent under the parity swap. However, these terms can only connect states with a different zero-mode occupation and are therefore not relevant for (\ref{eq:diag_equiv}) and subsequent use in the degenerate perturbative expansion about the special point. 

For all terms that {\em do} connect states with the same parity and band index we see that either (I) $\beta_1$ and $\beta_1^\dagger$ do not occur, or that (II) they both occur. For the case (I) where neither occur we see that for every contributing non-zero term in one sector there is an identical term with exactly the same coefficient in the other sector. For case (II) where both $\beta_1$ and $\beta_1^\dagger$  occur in order to calculate the corresponding term in the opposite sector we must parity swap {\em pairs} of coefficients (e.g. $U_{x_a,1} V_{x_b,1} \lr V_{x_a,1} U_{x_b,1} $) and therefore by \eqref{eq:UVdecomp2} they are equivalent provided $x_a$ and $x_b$ occur near each other.   

Let us now turn to our perturbation analysis.  This local restriction on $x_a$ and $x_b$ requires not only that $H_{\text{Perturb}}$ is local but also that $H_{\text{Perturb}}^s$ is local. For a finite system this limits the order at which  the perturbation expansions of even and odd sectors remain equivalent. This is not a problem however as we can make our  idealised wire as long as we wish.  A more problematic issue in this regard is showing the convergence of the perturbative series. 
For the groundstates $\ket{0}_e$ and $\ket{0}_o$  there are no problems and one finds the expected situation where the two states remain degenerate to an order of perturbation theory that is proportional to the system length $N_x$.  However, technical problems do arise when performing perturbative expansions of higher energy states which form part of the bulk. In this case, it becomes possible for divergences to occur when intermediate states in the expansion  have energies that are close to the energy of the state we are considering.

As alluded to above, to get around this problem we note that the special non-interacting parameters $\Delta=t$ and $\mu=0$ (see Ref. \onlinecite{Kitaev2001}) lie within the parameter space of the topological phase . In this case the Majorana  operators become completely localised around end points with $u_L(x)= \delta_{x,1}$ and $u_R(x)=\delta_{x,N_x}$. Furthermore we see that in this case the dispersion relation is flat and all states $\ket{j,N}_p$ with the same values of $N$ and $p$ are degenerate.  This simplifies things considerably because we can then use degenerate perturbative expansion \cite{Kato1949, Bloch1958,Messiah1961}  in $H_{\text{Perturb}}=H_{I}+H_\Delta+H_{\mu} $ where 
\bea
H_\Delta &=&( \Delta-t)  \sum_x c^\dagger_x c^\dagger_{x+1}+h.c. \\
H_\mu &=&  -\mu  \sum_x c^\dagger_x c_{x}
\eea
for each degenerate band. Since the general observation regarding matrix elements in the topological phase is also true at the special point we see that to order $s= N_x$ in any degenerate expansion, the matrix elements  $\phantom|_0\bra{j,N} H_{\text{Perturb}}^s \ket{k,N}_0$ and $\phantom|_1\bra{j,N} H_{\text{Perturb}}^s \ket{k,N}_1$  remain the same. As we are free to make the wire as long we like we can therefore say that, provided the perturbation expansion converges,  then $H_{(e)}$  and  $H_{(o)}$ are unitarily equivalent.

This is the main result of this work. An interesting special case of the above result is the half-interacting wire, i.e. a wire where the interactions only occur on the left or right hand side. From (\ref{eq:LRdecomp2}) we see that we can factor the even and odd parity sub-Hamiltonians  as
\bea
\label{eq:HeHo}
H_{(e)} &=& E+ S +L+R \\
H_{(o)} &=& E+ S +L-R \non .
\eea
where $E$ is the diagonal matrix containing all the original non-interacting energies $E_n$  and $S$ represents interacting terms that are the same in both even and odd sectors (i.e. those that do not contain any $\beta_1$ or $\beta_1^\dagger$ terms).   In this basis we see that if interactions appear only in the left of the system,  and not in the right of the wire,  then  $H^{(e)} = H^{(o)} $. In the appendix we analyse this particular scenario in more detail and outline its connection with previous works. \cite{Goldstein2012,Yang2014}

{\em Implications for the Majorana mode structure:}   For the non-interacting system, because we can move anti-symmetric considerations onto other quasi-particle operators $\beta^\dagger_n$ and  $\beta^\phd_n$, we can define
\bea
\gamma_R &=& \phantom i \sum \ket{n}_1  \phantom|_0 \bra{n} +\ket{n}_0  \phantom|_1 \bra{n} \\  
\gamma_L &=& i \sum \ket{n}_1  \phantom|_0 \bra{n} - \ket{n}_0  \phantom|_1 \bra{n} \non 
\eea
The unitary equivalence of the even and odd sectors means that we can proceed in a similar way when we allow $H_{I}$ to be non-zero. In principle we would like to write 
\bea
\bar{\gamma_R} &=& \phantom i \sum \ket{\bar{n}}_1  \phantom|_0 \bra{\bar{n}} +\ket{\bar{n}}_0  \phantom|_1 \bra{\bar{n}} \\  
\bar{\gamma_L} &=& i \sum \ket{\bar{n}}_1  \phantom|_0 \bra{\bar{n}} - \ket{\bar{n}}_0  \phantom|_1 \bra{\bar{n}} \non 
\eea
where  $ \ket{\bar{n}}_0  =  \ket{ \{0,\bar{n}\}} $  and $ \ket{\bar{n}}_1  = \ket{ \{1,\bar{n}\}}  $ and the integers $\bar{n}$, although they can no longer be related to the binary numbers indicating the occupancy of the other non-zero excitations in the non-interacting system,  still count the remaining degrees of freedom in the model.  However, in a practical calculation we would have obtained the eigenvectors $\ket{\bar{n}}_e$ and $\ket{\bar{n}}_e$  as opposed to $\ket{\bar{n}}_0$ and $\ket{\bar{n}}_1$ . As I mentioned above, it is only in the case of extremum energy states that the occupancy of the zero-mode is reliably inferred from the total parity. In addition to this we must also realise that in any numerical calculation the wave-functions will be returned with some arbitrary phase.  To solve this problem we can fix the relative phases of the even-odd wavefunctions using our bare-non interacting Majorana modes. For our situation with real coefficients only we calculate $
s_n^{(R)}=\text{sign} (\phantom|_o\bra{\bar{n}} \beta_1^\dagger + \beta_1 \ket{\bar{n}}_e)
$ and  set $\ket{\bar{n}}_o \rightarrow s^{(R)}_n \ket{\bar{n}}_o$ . Then, with 
$
s_n^{(L)}=\text{sign} (\phantom|_o\bra{\bar{n}} \beta_1^\dagger - \beta_1 \ket{\bar{n}}_e)
$, we can then write 
\bea
\label{eq:gamma_bar}
\bar{\gamma}_R &=& \phantom i \sum \;\;I \;\; \ket{\bar{n}}_o   \phantom|_e \bra{\bar{n}}  + \;\; I \;\; \ket{\bar{n}}_e  \phantom|_o \bra{\bar{n}}  \\
\bar{\gamma}_L &=& i \sum  s^{(L)}_n \ket{\bar{n}}_o  \phantom|_e \bra{\bar{n}}  - s^{(L)}_n \ket{\bar{n}}_e  \phantom|_o \bra{\bar{n}} .
\non 
\eea
We see that these operators behave as Majorana's should : $ \{  \bar{\gamma}_R, \bar{\gamma}_L  \} =0$ and $\bar{\gamma}^2 =I$ . 

The fact that the many-body Majorana operators are well defined  quasi-particle excitations has some interesting consequences when probing their structure using data obtained from the ground states of DMRG/MPS based variational techniques. \cite{Stoudenmire2011}   In this approach one first calculates both of the systems groundstates and then probes the cross correlators $ \phantom|_1 \bra{\bar{0}} \mathcal{O} \ket{\bar{0}}_0$. However it is clear from \eqref{eq:gamma_bar}  the many-body Majorana operators inherit their structure from all eigenstates of the system  and that position space structure of say $\bar{\gamma}_R$ could be very different from the single contributing  term $ \ket{\bar{0}}_o   \phantom|_e \bra{\bar{0}} + \ket{\bar{0}}_e   \phantom|_o \bra{\bar{0}} $.  Therefore simply probing  the ground state cross-correlators would not provide the full picture of the Majorana quasi-particle as we have come to understand it. This is in contrast to the linear Majorana operators obtained in the non-interacting limit, which can be defined using cross-correlators of any single even-odd pair of eigenstates.   There, when one examines the non-interacting states  $ \phantom|_1 \bra{0} \mathcal{O}_x \ket{0}_0$ with e.g. $\mathcal{O}_x = c_x^\dagger+ c_x$ one picks up the function $u_R(x)$ exactly. In the interacting case the correlator  $ \phantom|_1 \bra{\bar{0}} \mathcal{O}_x \ket{\bar{0}}_0$, while containing contributions from the linear terms, will also contain non-zero contributions from higher-order multinomials in the many-body Majorana expansion.\cite{gkells2014}   

In this paper I have argued that in the topological phase, even and odd parity-sectors remain equivalent despite the potential presence of local interaction  terms.  From this observation it follows that there are particle-hole symmetric many-body operators which connect states of even and odd parity at the same energy.  The  arguments given here, apply to all states of the system  and therefore  imply that the many-body Majorana modes are  true infinite-lifetime quasi-particles that are valid for the full Hilbert space. They therefore behave in much the same way as the linear Majorana operators of the non-interacting system. 

Although it is formulated specifically for the spinless p-wave model,  similar arguments should also apply to quasi-1-dimensional variants of the p-wave model \cite{ } and to models that obtain the p-wave symmetry through effective means (e.g. using combinations of Zeeman-splitting and spin-orbit and proximity coupling) \cite{} .  The reason for this is that, regardless of the precise underlying mechanism, the mean-free descriptions of the associated topological phases contain Majorana bound states with the same particle-hole structure as \eqref{eq:UVdecomp}, but where the $x$-index now represents additional position and internal indices.  Therefore,  provided the interacting term only connects local position indices, the argument describing the matrix elements should follow through in the same way.  

One caveat is that for these more general models  we cannot simply perturb away from a special point with universally flat dispersion. Therefore in order to show that these expansions converge, a more sophisticated resolvent treatment of the nearly degenerate bands would be needed.  However, it is important to emphasise that, in these cases, the issue is not whether the perturbative expansions in the even and odd sectors are different,  but rather showing that both expansions actually converge.

{\bf Acknowledgements} I  thank Niall Moran, Awadhesh Narayan, Jiri Vala, Joost Slingerland, Dganit Meidan, Alessandro Romito, Emil Bergholtz, Piet Brouwer, Smitha Vishveshwara, and Diptiman Sen for fruitful discussions regarding this topic. This work was partially supported by the Science Foundation Ireland award 10/IN.1/I3013 .

\appendix

\section*{Appendices}

In these Appendices I present  numerical results that back up the main claims of the manuscript, details on how to calculate arbitrary matrix elements in the noninteracting basis, and a further discussion of the situation with interactions in only one half of the wire.

\section{Numerical results}

\begin{figure}[!]
\includegraphics[width=.4\textwidth,height=0.30\textwidth]{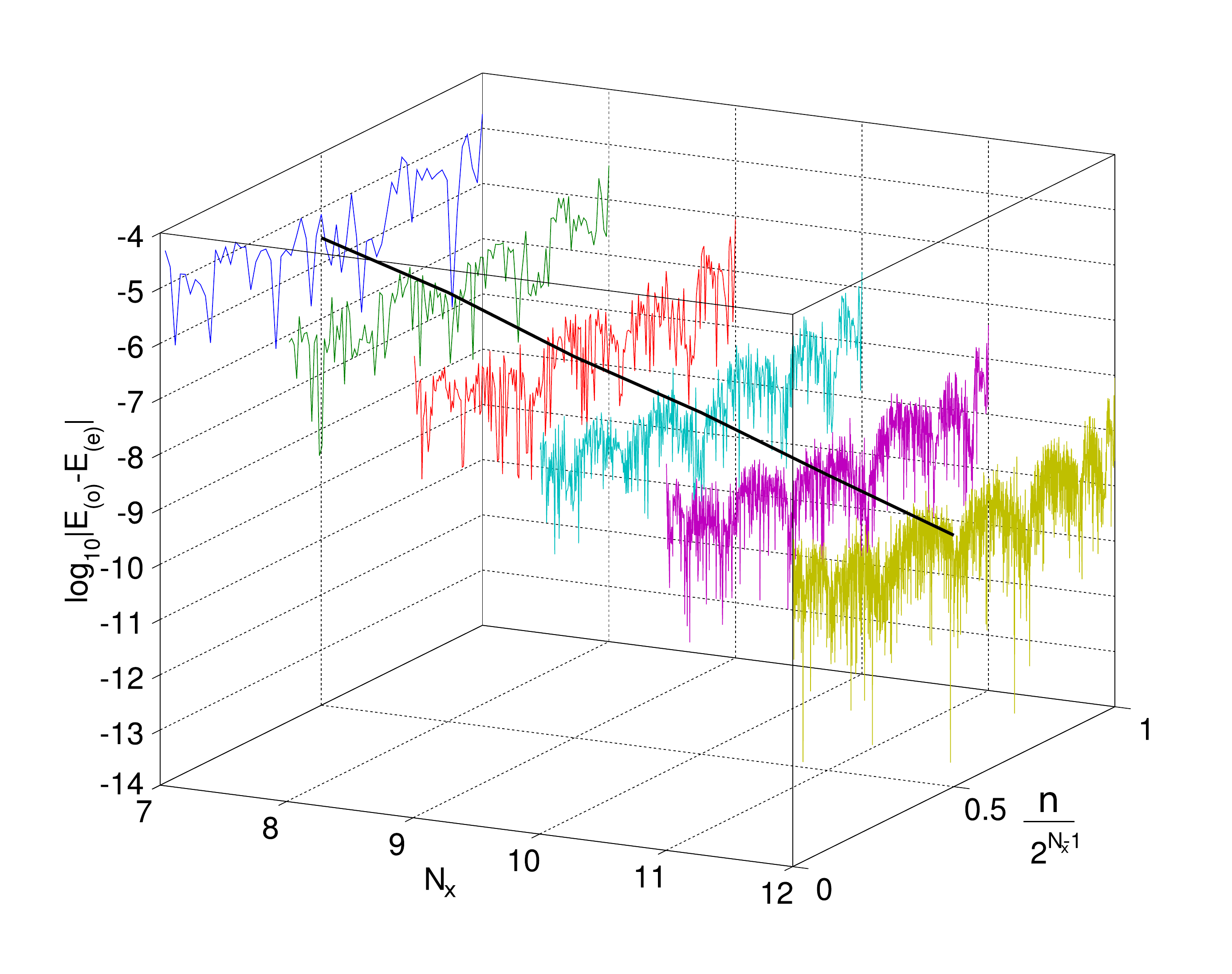}
       \caption{\label{fig:exp1}  The difference in energy between eigenpairs of different parity decreases exponentially with system size. The black line runs through the mean energy difference. The parameters used for this plot are $t=1$, $\Delta=0.98$, $\mu=-0.02$ and $V_{\text{int}}=0.3$. The parameters are chosen so that the Majorana bound states are closely confined to the system edges.   } 
\end{figure}

Figure \ref{fig:exp1} shows the difference in energy between all parity eigenpairs at different lengths. The key feature to notice is that the energy difference between all eigenstates decays exponentially with the system length.  This corroborates the main claim of this paper, albeit with very small system sizes. 

\section{Matrix elements in the non-interacting eigenbasis}
In the main text the eigenbasis of the non-interacting system is used to construct matrix elements of the full interacting Hamiltonian. In this appendix,  we review these calculations, following the notation of Ref. \onlinecite{Ring2004}.  Starting with the general form of quadratic Hamiltonian
\be
H_0 = \frac{1}{2} \left[\begin{array}{cc} c^\dagger_{\lr} \quad c^{\phantom \dagger}_\lr \end{array}  \right]  \left[\begin{array}{cc} f & g \\ g*& -f^T \end{array}  \right]  \left[\begin{array}{c} c^\phd_{\ud}  \\ c^{ \dagger}_\ud \end{array}  \right] 
\ee
where 
\be
\left[\begin{array}{cc} c^\dagger_{\lr} \quad c^{\phantom \dagger}_\lr \end{array}  \right] = \left[\begin{array}{cc} c^\dagger_1 ,  c^\dagger_2 , ... c^\dagger_N,  c^{\phantom \dagger}_1,  c^{\phantom \dagger}_2 ,... c^{\phantom \dagger}_N  \end{array}  \right] 
\ee
The system may be cast in terms of free fermions using  a Bogoliubov transformation  
\bea
&&\left[\begin{array}{cc}\beta_1^\dagger,...,\beta_N^\dagger,  & \beta_1,...,\beta_N \end{array}  \right] \\&& =  \left[\begin{array}{cc} c^\dagger_{\lr} \quad c^{\phantom \dagger}_\lr \end{array}  \right]  \left[\begin{array}{cc} U & V^* \\ V & U^* \end{array}  \right] = [\psi^\dagger_{\lr} ] [W] .
\eea
It is useful to introduce 
\bea
\rho^n_{xx'} &=& \bra{n} c_x^\dagger c_{x'}^{\phd} \ket{n} \non \\
\kappa^n_{xx'}&=& \bra{n} c_x^\phd c_{x'}^{\phd} \ket{n}
\eea
In terms of $U$ and $V$ matrices we may write :
\be
\rho = V^* V^T \quad \kappa = V^* U^T.
\ee
The general form of the interaction term can be written as
\be
H_{I} = \frac{1}{4}  \sum_{x_1 x_2 x_3 x_4}  \bar{\nu}_{x_1 x_2 x_3 x_4}  c^\dagger_{x_1}  c^\dagger_{x_2}  c^\phd_{x_4}  c^\phd_{x_3}  
\ee
where we use the standard convention $\bar{\nu}_{x_1 x_2 x_3 x_4} = \nu_{x_1 x_2 x_3 x_4}-\nu_{x_1 x_2 x_4 x_3}$.   Using the eigenbasis of the original Hamilton we expand out the terms in the interaction term as
\bea
H_{I} =  \frac{1}{4}   \sum_{x_1 x_2 x_3 x_4} \bar{\nu}_{x_1 x_2 x_3 x_4}   &&  \sum_i (U_{x_{1} i}^* \beta_i^\dagger + V_{x_{1} i}^{\phantom*}  \beta_i^{\phd}) \\
&& \sum_j (U_{x_{2} j}^* \beta_j^\dagger + V_{x_{2} j}^{\phantom*}  \beta_j{\phd} ) \non \\ 
&& \sum_k (U_{x_{3} k}^{\phantom*} \beta_k^{\phd }+ V_{x_{3} k}^* \beta_k^\dagger )  \non \\
&&\sum_l(U_{x_{4} l}^{\phantom*} \beta_l^{\phd} + V_{x_{4} l}^* \beta_l^\dagger ) \non.
\label{eq:H_expand}
\eea
In the case of the p-wave wire in the main text we set $\nu_{x_1 x_2 x_3 x_4} = I_{\text{int}} (x_1,x_2) \delta_{x_1,x_3} \delta_{x_2,x_4}$ with $x_2 \ne x_1$. 

Expanding out the the full Hamiltonian $H=H_0 + H_{\text{int}}$ we have
\bea
H&=&H^0 + \sum_{k_1 k_2} H^{11}_{k_1 k_2} \beta^\dagger_{k_1} \beta^\phd_{k_2}  \\ &+& \non  \frac{1}{2} \sum_{k_1,k_2} (H_{k_1 k_2}^{20} \beta^\dagger_{k_1} \beta^\dagger_{k_2} +\text{h.c.}) \\
\non &+& \sum_{k_1 k_2 k_3 k_4}  ( H^{40}_{k_1 k_2 k_3 k_4} \beta^\dagger_{k_1}  \beta^\dagger_{k_2}  \beta^\dagger_{k_3}  \beta^\dagger_{k_4 } +  \text{h.c.}) ) \\
\non &+&\sum_{k_1 k_2 k_3 k_4}  ( H^{31}_{k_1 k_2 k_3 k_4} \beta^\dagger_{k_1}  \beta^\dagger_{k_2}  \beta^\dagger_{k_3}  \beta^\phd_{k_4 } +  \text{h.c.}) ) \\
\non &+&\frac{1}{4} \sum_{k_1 k_2 k_3 k_4}  ( H^{22}_{k_1 k_2 k_3 k_4} \beta^\dagger_{k_1}  \beta^\dagger_{k_2}  \beta^\phd_{k_4}  \beta^\phd_{k_3} +  \text{h.c.}) )
\eea
where if we set 
\bea
h&=&f+\Gamma \non  \\
\Gamma_{lm} &=& \sum_{pq} \bar{\nu}_{lqmp} \rho_{pq} \non \\ 
\Delta_{lm} &=& \frac{1}{2} \bar{\nu}_{lqmp} \kappa_{pq} \non \\
F^0 &=&  Tr( f \rho) - \frac{1}{2} Tr(g \kappa^*+g^* \kappa) \non
\eea
we can write
\bea
H^0 &=& F^0 +\frac{1}{2} Tr(\Gamma^*\rho-\Delta^*\kappa), \non \\
H^{11} &=& U^\dagger h U^{\phantom*} - V^\dagger h^T V^{\phantom*}  +U^\dagger \Delta V^{\phantom*}  - V^\dagger \Delta^* U^{\phantom*} , \non \\
H^{20} &=& U^\dagger h V^* - V^\dagger h^T U^* +U^\dagger \Delta U^* - V^\dagger \Delta^* V^*, \non
\eea
 and
 \bea
 H^{40}_{k_1 k_2 k_3 k_4} &=& \frac{1}{4} \sum_{x_1 x_2 x_3 x_4} \bar{\nu}_{x_1 x_2 x_3 x_4} U^*_{x_1 k_1} U^*_{x_2 k_2} V^*_{x_4 k_3} V^*_{x_3 k_4} \non  \\
  H^{31}_{k_1 k_2 k_3 k_4} &=& \frac{1}{2} \sum_{x_1 x_2 x_3 x_4} \bar{\nu}_{x_1 x_2 x_3 x_4} (U^*_{x_1 k_1} V^*_{x_4 k_2} V^*_{x_3 k_3} V^{\phantom*}_{x_2 k_4} \non \\ &+& V^*_{x_3 k_1} U^*_{x_2 k_2} U^*_{x_1 k_3} U_{x_4 k_4}) \non \\
  H^{22}_{k_1 k_2 k_3 k_4} &=& \sum_{x_1 x_2 x_3 x_4} \bar{\nu}_{x_1 x_2 x_3 x_4} \times \non \\ && [U^*_{x_1 k_1} V^*_{x_4 k_2} V_{x_2 k_3} U^{\phantom*}_{x_3 k_4}) \\  &-& U^*_{x_1 k_2} V^*_{x_4 k_1} V_{x_2 k_3} U^{\phantom*}_{x_3 k_4})]   \non \\
   &+&  [U^*_{x_1 k_1} U^*_{x_2 k_2} U_{x_3 k_3} U_{x_4 k_4}   -(k_3 \lr k_4)   \non \\ &+& V^*_{x_3 k_1} V^*_{x_4 k_2} V_{x_1 k_3} V_{x_2 k_4} ]  \non 
 \eea

\vspace{2mm} 
A critical feature of the main text is that we distinguish between terms in $H^s = (H_0 +H_I)^s$ that 
\begin{enumerate}
\item   don't include either $\beta^\dagger_1$ or $\beta^{\phd}_1$  (we call these terms S) 
\item   include either $\beta^\dagger_1$ or $\beta^{\phd}_1$  (we denote these terms $D$ )
\item   include  $\beta_1^\dagger$'s and $\beta_1$'s in equal number (we call these $B$)
\end{enumerate}
Crucial to this line of reasoning will be the observation  that, in  Eq. (\ref{eq:H_expand}),  for every occurrence of $\beta_1^\dagger$  with coefficient $U_{x_1,1}$ (or $ V_{x_1,1}$) there is a parity swapped occurrence of  $\beta_1^\phd$ with coefficient $V_{x_1,1}$ (or $U_{x_1,1}$ ) coming from the same term in the expansion. Furthermore, this parity swapped contribution must act in the opposite parity sector. For those terms S that do not contain either $\beta^\dagger_1$ or $\beta^{\phd}_1$ it is a trivial task to show that matrix elements  are the same in each sector.  For those terms $D$  that contain {\it either} $\beta_1^\dagger $ {\it or} $\beta_1$, using Eq. (10) in the main text, we see that matrix elements can be sub-divided into contributions from the left and right of the wire
 \bea
\label{eq:LRdecomp1}
\phantom|_e \bra{n} D_e  \ket{m}_e  = L_{nm} + R_{nm} \\
\phantom|_o \bra{n} D_o  \ket{m}_o  = L_{nm} - R_{nm}\non
\eea
where $D_e$ is the same as $D_o$ but with the swap $\beta_1 \lr \beta_1^\dagger$  and corresponding coefficient swap $U_{x,1} \lr V_{x,1}$ at the same value of $x$. However, while these terms are different in even and odd sectors, it is clear that they cannot connect states that share both the same band ($N$) and parity $(p)$  index, i.e. if $\ket{n}=\ket{j,N}$, $\ket{m}=\ket{k,N}$ and $p$ denotes the occupation of the zero mode ($0$ or $1$) then
\bea
\label{eq:LRdecomp3}
\phantom|_p \bra{j,N} D  \ket{k,M}_p  = (L_{nm} \pm (-1)^p R_{nm}) [1-\delta_{NM}] . \non
\eea

For the terms $B$ where both $\beta^\dagger_1$ and $\beta^\phd_1$ occur we see that we can only connect matrix elements of the same band. 
\be
\phantom|_p \bra{j,N} B \ket{k,M}_p  = B^{(p)}_{nm} \delta_{NM}
\ee
The essential question is then in what scenario does $B^{(0)}_{nm}=B^{(1)}_{nm}$. Our parity swapping arguments imply that if a matrix element $ B^{(0)}_{nm} \ne 0$ then there will also exist another non-zero matrix element in the other sector $B^{(1)}_{nm}$  whose value is related  $ B^{(0)}_{nm}$  by the swap $U_{x,1} \lr - V_{x,1}$ or $V_{x,1} \lr - U_{x,1}$ in {\em each} of the contributing coefficients. 

Now this is where the localised nature of the Majorana modes and the Hamiltonian comes into play.  If $H_I$ is a local such that $|x_1 -x_2|  < l $ for some finite length $l$ , then $H_l^s$ will  only will have non-zero elements between between sites within a distance $s l$).  Considering then elements $\bra{n}H^s\ket{m}$  we see clearly that as long as $s$ is not comparable to the system size $N_x$, then the x-indices of the coefficients occurring in the expansion  of  $H_I^s$ cannot occur at opposite sides of the system. (i.e. $x_1$ \& $x_2 \in x_L$ or $x_1$ \& $x_2 \in x_R$). In this case, because of relations  Eq. (10)  of the main text, one always has equality between coefficient pairs (e.g. $U_{x_{L},1} U_{x_{L},1} $ , $U_{x_R1} V_{x_R,1} $ .... ) and their parity swapped counterparts (e.g. $V_{x_{L},1} V_{x_{L},1} $ , $V_{x_R,1} U_{x_R,1} $ ....).  This implies that for matrix elements between states within the same band,  every contribution in the even sector has an equal counterpart in the odd sector.

\section{ The half-interacting wire} In the eigen-basis of the non-interacting system the non-interacting Hamiltonian is diagonal
\be
H=  \left[\begin{array}{cc} H_{(e)}& 0 \\ 0 & H_{(o)} \end{array}  \right] = \left[\begin{array}{cc} E_{(e)}& 0 \\ 0 & E_{(o)} \end{array}  \right] 
\ee
where, if there are zero modes the diagonal matrices $E_{(e)} =E_{(o)}$.  The zero-mode operators in this basis are defined as
\bea
\beta^\dagger_1&=& \sum \ket{ \{1,n_2,n_3,...\}} \bra{ \{ 0, n_2, n_3,...  \}}  \\
\beta^\phd_1&=& \sum \ket{ \{0,n_2,n_3,...\}} \bra{ \{ 1, n_2, n_3,...  \}} \non
\eea
which as matrices take the form
\be
\beta_1^\dagger = \left[\begin{array}{cc} 0 &  N_{e} \\  N_{o} & 0 \end{array}  \right],  \quad \beta_1^\phd = \left[\begin{array}{cc} 0 & N_{o}  \\  N_{e}  & 0 \end{array}  \right]
\ee
where the sub matrices $N_{e/o}$ are diagonal with elements 1 or 0 depending on the occupation of the $n_1$ zero mode in that sector.  For example  in this notation $N_e$ has a $1$ on the diagonal if the $n_1=1$ in $\ket{ \{n_1,n_2,...\}}_e$. In the diagonal basis we have  $N_e = I-N_o$ and we could for example  write
\be
\beta_1^\dagger=\left[\begin{array}{cc} 0 &  N_{e} \\   I-N_{e} & 0 \end{array}  \right],  \quad
\beta_1^\phd=\left[\begin{array}{cc} 0 & I-N_{e}  \\  N_{e}  & 0 \end{array}  \right]
\ee
In this case then the Majorana operators are 
\bea
\label{eq:gamma_def}
\gamma_R &:=& ~ (\beta_1^\dagger +\beta_1^\phd) = \sigma^x \otimes I \\
\gamma_L &:=& i (\beta_1^\dagger -\beta_1^\phd) = \sigma^y \otimes F \non
\eea
where the diagonal operator $F=I-2N_e = -I+2N_o$. Both operators $\gamma_R$ and $\gamma_L$ take an eigenstate in one sector, (which in this basis are column vectors with one element 1 and all others 0 : i.e. $\ket{n} = [0000...1...0000]^T$),  to the corresponding parity swapped state  in the other sector.  With our convention $\gamma_R$ does not introduce a phase shift or change of sign. On the other hand $\gamma_L$, upon swapping the state to the other sector, introduces a $\pm i$ phase shift. 

In the main text we point out that for the interacting system, when the bare system has well-seperated Majorana zero-modes,  the sub-matrices of $H_{\text{I}}$ that do not connect basis elements differing in the occupation of $n_1$,  (i.e. represent terms in the expansion of $H_{\text{I}}$ that contain  $\beta_1^\dagger$ or $\beta_1$) are the same for even and odd sectors and we can denote them as $S$.  For terms that do contain $\beta_1^\dagger$ or $\beta_1$ we can decompose them into the left and right contributions:
\bea
\label{eq:LRdecomp}
\phantom|_e \bra{n} D_{(e)} \ket{m}_e  &=&L_{nm}+R_{nm} \\
\phantom|_o \bra{n} D_{(o)} \ket{m}_o &=& L_{nm}-R_{nm} \non
\eea
and see that generally $\phantom|_e \bra{n} H_{I} \ket{m}_e \ne  \phantom|_o \bra{n} H_{I} \ \ket{m}_o$. The full structure of the even and odd parity sub-Hamiltonians can thus be written as
\bea
\label{eq:HeHo}
H_{(e)} &=& E+ S +D_{(e)} \\
H_{(o)} &=& E+ S +D_{(o)} \non .
\eea
where $E$ is the diagonal matrix containing all the original non-interacting energies $E_n$  and we again note that the interaction parameter $I$ is contained within $S$  and $D$ matrices.  From  here it is easy  to see that if interactions appear only in the bulk and left of the system,  {\em  not in the right of the wire},  then  $H^{(e)} = H^{(o)} $. With some trivial changes of sign conventions we can make an identical argument for the system when only the right-hand side is interacting.  

It is interesting to see how the above position dependent interaction terms will affect the structure of the Majorana modes in the non-interacting eigenbasis.  Intuitively we know what should happen: since the interactions have no effect on the RHS of the system  we expect  that the $\gamma_R $ should remain the same while $\gamma_L$ should change.  

Lets formulate how this happens.  As $H^e =H^o$ then we know that in this basis the new eigenstates will look the same in the even or odd sectors. More precisely if $\bar{\ket{n}}$ are the new eigenstates of the system, then each  $_e\braket{m}{\bar{n}}_e$ will have a counterpart $_o \braket{m}{\bar{n}}_o$ with the same value in the other sector. Hence the operator $\gamma_R$  which is the original (non-interacting) right-hand-side Majorana, will continue to map between even-odd eigenstates of the interacting Hamiltonian.  
\be
{\gamma_R}\ket{\bar{n}}_{e} = \ket{\bar{n}}_{o}, \quad {\gamma_R}\ket{\bar{n}}_{o} = \ket{\bar{n}}_{e}
\ee
or 
\bea
\bar{\gamma_R}&=&  (\bar{\beta}_1^\dagger +\bar{\beta}_1^\phd) =(\beta_1^\dagger +\beta_1^\phd)=\gamma_R .
\eea
 On the other hand, although the sum $(\beta_1^\dagger+\beta_1^\phd)$ is invariant under this position dependent interacting term,  the individual operators $\beta_1$ ,  $\beta_1^\dagger$ and  thus $\gamma_L$ are not.  We can formulate this as
\be
\bar{\beta_1}^\dagger = \left[\begin{array}{cc} 0 & M \\ I-M & 0 \end{array}  \right], \quad \bar{\beta_1} = \left[\begin{array}{cc} 0 & I-M \\ M & 0 \end{array}  \right] 
\ee
where $M$ is a symmetric matrix . It immediately follows that in the non-interacting eigenbasis
\bea
\bar{\gamma_L} &:=& i (\bar{\beta}_1^\dagger -\bar{\beta}_1^\phd)  = i \left[\begin{array}{cc} 0 & -I+2M \\ I-2M & 0 \end{array}  \right] \\ &=&
 i \left[\begin{array}{cc} 0 & -A \\ A & 0 \end{array}  \right] = \sigma^y \otimes A
\eea
From the Majorana condition $\gamma_L^2 =I$ we immediately see that $R^2 =I$ and thus that $M^2=M$. Since  $M$ is symmetric and idempotent it is by definition an orthogonal projector.  In the non-interacting limit we see that $A=F$ and we retain our original expressions \eqref{eq:gamma_def} for $\gamma_L$ and $\gamma_R$.

Using the expressions for $\bar{\beta_{1}^\dagger} $ and $\bar{\beta_{1}}^\phd$ we can calculate the density operator 
\be
\bar{\rho_1}=\bar{\beta_{1}^\dagger} \bar{ \beta_1} =  \left[\begin{array}{cc} M & 0 \\ 0 &I-M  \end{array}  \right]
\ee
where we have used the fact that $M^2 =M$.   In the non-interacting limit ( where $\bar{n}=n$ ) $M$ is diagonal with a $1$ or a $0$ depending on whether the states $\ket{n}_e$ have the fermionic mode occupied or empty.  We see then, that in this limit, $M=N_e$. 

One of the compelling features of the half-interacting wire is that it allows us to see simply why the local nature of the interactions are essential for preserving the full topological degeneracy. It is tempting to try to construct a general argument inferring the existence of the Majorana in the interacting region from the fact that our Majorana on the RHS is the same as before.  The technical problem with this argument is that  even if we allow arbitrarily high interaction strengths on the LHS , we cannot not destroy this precise $H^{(e)}=H^{(o)}$ simply because the left-hand side Majorana mode can always shift position and decay into the bulk of the non-interacting region. 



\begin{thebibliography}{}
\bibitem{Continuum}These parameters can be related to the continuum parameters through  $\mu=\mu_0-2t$, $t=1/(2 m a^2)$  and $\Delta_n = \Delta_n/(2a)$, where $a$ is the lattice constant and the wire length $L=(N_L+1)a$.
\end{thebibliography}
\end{document}